\documentstyle[preprint,aps,epsf,titlepage]{revtex}
\begin{document}
\tightenlines
\title{Looking Forward to Pricing Options from Binomial Trees}
\author{\it{ Dario Villani$^{a}$} and \it{Andrei~E.~Ruckenstein$^{b}$}}
\address{(a) 10 Brookside Drive, Greenwich, CT 06830 \\
(b) Department of Physics and Astronomy,
Rutgers University,\\
136 Frelinghuysen Road, Piscataway, NJ 08854}
\date{\today}
\maketitle

\begin{abstract}
We reconsider the valuation of barrier options by means of binomial
trees from a ``forward looking" prospective rather than the more
conventional ``backward induction" one used by standard approaches.
This reformulation allows us to write closed-form
expressions for the value of European and American put
barrier-options on a non-dividend-paying stock.
\end{abstract}

\newpage

\section{Introduction}

Options are financial contracts which give the holder the right to buy
(call options) or sell (put options) commodities or securities for a
predetermined exercise (or strike) price by a
certain expiration date~\cite{hull}.
Conventional European (American) options can be exercised only on (at any
time up to) the expiration date. Since the option confers on its holder a
right with no obligation, it should carry a price at the time of
contract. It is  the classic work of Black, Scholes
and Merton~\cite{black-scholes,merton} which suggested a strategy for
determining a fair price for the option in a risk-free environment.

Closed-form valuation within the Black-Scholes-Merton equilibrium pricing
theory~\cite{black-scholes,merton} is only possible for a small subset of
financial derivatives. In the majority of cases one must appeal to
numerical techniques such as Monte Carlo simulations, or finite
difference methods and much of the effort in the field has been in
developing efficient algorithms for numerically solving the Black-Scholes
equation~\cite{hull}. An alternative direction has
been the evaluation of discrete-time, discrete-state stochastic
models of the market on binomial and trinomial trees~\cite{crr,cox}.
Not only is this discrete approach intuitive and easily accessible to
a less mathematically sophisticated audience; but
it also seems to us to be a more accurate description of market dynamics
and better suited
for evaluating more involved
financial instruments. Moreover, the few exact Black-Scholes results
available can be recovered in the appropriate continuous-time trading
limit. The main difficulty in pricing with binomial trees has been the
non-monotonic numerical convergence and the dramatic increase
in computational effort with increasing
number of time steps~\cite{derman,figlewski}. For example, the state of
the art calculations involve memory storage scaling
linearly (quadratically) with the number of time steps, $N$, for European
(American) options, while the computation time increases like $N^2$ in
both cases~\cite{wilmott}.

In this paper we reconsider valuation on binomial trees from what we call
a ``forward looking" prospective: we imagine acting as well-educated
consumers who attempt to eliminate risk and estimate the future
expected value of an option according to some reasonable dynamical model.
We will regard the movement of the price on the
tree as a random walk (with statistical properties consistent with a
risk-neutral world) with ``walls" imposed by the nature of the option,
such as the possibility of early exercise (American options) or the
presence of barriers. The resulting mathematical formulation then has two
conceptually distinct components: the first ingredient is an explicit
description of the possible ``walls". For example, in the case
of barrier American options both the barrier and the ``early exercise"
surface need to be specified. The second step will be to compute the
probability that the price reaches particular values at every accessible
point on the tree. This involves counting the number of paths reaching
that point in the presence of ``walls", a somewhat involved but
exactly solvable combinatorics problem. Once these two steps (specifying
the walls and computing the probabilities) are accomplished the value of
both European and American options, with and without barriers, can be
written down explicitly. In an attempt to be pedagogical, we will
limit ourselves to the simplest put options: European,
simple American and European with a straight ``up-and-out"
barrier. Although the calculation can be simply extended to the barrier
American option that discussion merits a separate publication.

As far as we know, in the case of trees explicit formulas
like the ones we are proposing exist in the literature only in the
simplest case of conventional European options~\cite{cox,pliska}.
For the more complicated case of American options, the main issues are
best summarized in the last chapter of Neil Chriss' book~\cite{chriss}:
``The true difficulty in pricing American options is determining exactly
what the early exercise boundary looks like. If we could know this {\it a
priori} for any option (e.g., by some sort of formula), we could produce
pricing formulas for American options." Below we propose a solution to
this problem in the context of binomial trees. Our
formulation complements the earlier studies of American options in the
limit of continuous-time trading~\cite{geske,kim,gao} which also focus on
the presence of an early exercise boundary for the valuation of
path-dependent instruments. The study of the continuum limit of our
formulas is instructive and will be left for a future publication.

\section{Binomial Trees}

To establish notation we begin by dividing the life of an option,
$T$, into $N$ time intervals of equal length, $\tau = T/N$. We assume
that at each discrete time $t_i = i\tau$ ($i=0,1,2,...,N$) the stock
price moves from its initial value, $S_0 = S(t_0 =0)$, to one of two
new values: either up to $S_0 u$ ($u > 1$) or down to $S_0 d$
($d < 1$)~\cite{wilmott}. This process defines a tree with nodes
labeled by a two dimensional vector, $(i,j)$ ($i=0,1,2,...,N;
j=0,1,...,i$) and characterized by a stock price
$S(i,j) = S_0 u^j d ^{i-j}$, the price reached at time $t_i =
i\tau$ after $j$ up and
$i-j$ down movements, starting from the original price $S_0$.
The probability of an up (down) movement will be denoted by
$p_u$ ($p_d = 1-p_u$); and thus each
point on the tree is also characterized by the probability,
$p_u ^j (1-p_u)^{i-j}$,
which represents the probability associated with a single path of $i$
time steps, $j$ ($i-j$) of which involve an increase (decrease) in the
stock price. Computing the
probability of connecting the origin with point $(i,j)$ requires,
in addition to the single path probability, a
factor counting {\em the number} of such possible paths in the
presence of a barrier and/or the possibility of early exercise. The
calculation of this degeneracy factor involves the details of each
financial derivative and it will be discussed in turn for each of our
examples.

The binomial tree model introduces three free parameters,
$u, d$ and $p_u$. Two of these are usually fixed by requiring that
the important statistical properties of the random
process defined above, such as the mean and variance, coincide with those
of the continuum Black-Scholes-Merton theory~\cite{hull}. In particular,
\begin{eqnarray}
\label{statprop1}
p_u u~+~\left (1-p_u \right ) d~&=&~e^{r\tau}\\
\label{statprop2}
e^{r\tau}\left (u~+d\right )~-~ud~-e^{2r\tau}~&=&~\sigma ^2 \tau,
\end{eqnarray}
where $r$ is the risk-free
interest rate,
and the volatility, $\sigma$, is a measure of the variance of the stock
price. We are left with one free parameter
which can be chosen to simplify the theoretical analysis;
one might choose, for example,
$u=1/d$~\cite{crr}, which simplifies the tree geometry by
arranging that an up motion followed by a down motion leads to no change
in the stock price. This condition together with
(\ref{statprop1}) and (\ref{statprop2}) implies:
\begin{eqnarray}
\label{parameters1}
u~&=&~e^{\sigma \sqrt{\tau}}\\
\label{parameters2}
d~&=&~e^{-\sigma \sqrt{\tau}}\\
\label{parameters3}
p_u~&=&~\frac{e^{r\tau}~-~d}{u~-~d}.
\end{eqnarray}
We stress that Equations (1-5)
are to be regarded as short-time approximations where terms higher order
in $\tau$ were ignored.

With these definitions out of the way we can begin discussing the
valuation of put options with strike price $X$ and expiration time
$T$.

\subsection{European Put Options}

The simple European put option is a good illustration of our ``forward
looking" approach. We are interested in all those paths on the tree
which, at expiration time $i=N$, reach a price, $S(N,j)=S_0 u^j
d^{N-j} < X$, for which the option should be exercised. That
implies that $j\leq j^* = \mbox{Int}\: [\ln (X/S_0 d^N)/ \ln (u/d)]$,
where $\mbox{Int}$
refers to the integer part of the quantity in square brackets. The
mean value of the option at expiration can then be written as a sum over
all values of
$j\leq j^*$ of the payoff at $j$, $X-S_0 u^j d^{N-j}$, multiplied by the
probability of realizing the price $S(N,j)=S_0 u^j d^{N-j}$ after $N$
time steps,
$P[N,j]$. As already mentioned above,
$P[N,j] =\aleph _E [N,j] p_u ^j (1-p_u )^{N-j}$, where $\aleph _E [N,j]$
counts the number of paths starting at the origin and reaching
the price $S(N,j)$ in $N$ time steps. For
the case of conventional European options this is just the number of paths
of $N$ time steps, with $j$ up and $N-j$ down movements of
the price, and is thus given by the binomial coefficient,
\begin{equation}
\label{alephev}
\aleph _E [N,j] ~ = ~\left (\begin{array}{c}
N\\j
\end{array}
\right )~=~\frac{N!}{j!(N-j)!}.
\end{equation}

The resulting expression for the
mean value of the option at maturity is then discounted to the time of
contract by the risk-free interest rate factor,
$e^{-rT}$, to determine the
current expected value of the option:
\begin{equation}
\label{valueev}
\bar{V} _E = e^{-rT}\sum _{j=0} ^{j^*} \left (\begin{array}{c}
N\\j
\end{array}
\right )
p_u ^j (1-p_u)^{N-j}
\left (X-S_0 u^j d^{N-j}\right ).
\end{equation}
This expression is not new: it was first discussed by Cox and
Rubinstein~\cite{cox} who also showed that in the appropriate
continuous trading-time limit ($\tau \rightarrow 0$) (\ref{valueev})
reduces to the Black-Scholes result~\cite{black-scholes}.

\subsection{European Put Barrier Options}

We are now ready to extend (\ref{valueev}) into an exact formula for
the mean value of an European put option with a barrier. Although our
approach can be used for other barrier instruments, we consider the
simplest case of an ``up-and-out" put option which ceases to exist when
some barrier price,
$H > S_0$, higher than the current stock is reached.
With the choice $u=1/d$ an
explicit equation for the nodes of the tree which constitute the barrier
can be written down:
\begin{equation}
\label{barrier}
S(j_B +1 +2h, j_B + 1 +h) = S_0 u^{j_B + 1 + h} d^h, ~~~~~h=0,1,...,h_B
\end{equation} Here,
$j_B =  \mbox{Int}\left [ \ln\left (H/S_0\right) /\ln (u)\right]$ defines
the first point just above the barrier, $(j_B +1 , j_B +1)$, and $h_B$
labels the last relevant point on the barrier corresponding to the
time closest to the maturity of the option, i.e.,
$h_B = \mbox{Int}\left[(N-j_B -1)/2\right ]$.

Since the probability that any allowed path starting with
the present stock price, $S_0$, reaches an exercise price at maturity,
$S(N,j) < X$, is still
$p_u ^j (1-p_u )^{N-j}$ (with
$j \leq j^*$) the average value of the European barrier option can be
written in a form similar to (\ref{valueev}):
\begin{equation}
\label{valueeb}
\bar{V}_{EB} = e^{-rT}\sum _{j=0} ^{j^*} \aleph _{EB} [N,j]~
p_u ^j (1-p_u)^{N-j}
\left (X-S_0 u^j d^{N-j}\right ),
\end{equation}
where $\aleph _{EB} [N,j]$ is the number of paths $N$ time-steps long
involving $j$ up and $N-j$ down movements of the price excluding
those paths reaching any of the points on or above the barrier
(\ref{barrier}). As we will explain below,
$\aleph _{EB} [N,j]$ is given by
\begin{equation}
\label{excluded}
\aleph_{EB} [N,j] ~=~ \left(\begin{array}{c}
N\\j
\end{array} \right )~-~\sum _{h=0} ^{h_M} \aleph
_{EB} ^{res}[j_B +1+2h, j_B + 1+h]
\left (
\begin{array}{c}
N - j_B -1 -2h\\
j-j_B -1 -h
\end{array}\right ),
\end{equation}
where the second term on the right-hand side represents the contribution
from the unwanted paths which hit the barrier (\ref{barrier}) before
reaching an exercise point $(N,j)$.

To understand the form of the excluded contribution in
(\ref{excluded}) we first note that reaching the
excluded region requires that the path hits the barrier at least once.
One might think that the number of unwanted paths can then be
calculated by (i) counting the number of paths connecting the origin to
a given point on the barrier; (ii) multiplying this by the number of
paths connecting that point on the barrier with the exercise point
$(N,j)$~\cite{distance} -- this includes all paths which wander {\em into}
the above-barrier region; and finally (iii) summing over all
points of the barrier (\ref{barrier}). However, a particular path
reaching a given point on the barrier might have already hit any of
the previous barrier points, and thus it would also be counted in the
contribution in (ii) from all paths starting at the first barrier point
reached by the particular path under consideration. Thus, summing
indiscriminately over barrier points would
lead to overcounting unless, in (i), we only include those paths which
hit the barrier for the first time. In other words, (i) must only include
paths starting from the origin which reach the particular point on the
barrier without having previously visited any other barrier point. The
number of such restricted paths (reaching the point
$(j_B +1 +2h, j_B +1 +h)$) is what we denoted by
$\aleph _{EB} ^{res}[j_B +1+2h, j_B + 1+h]$ in (\ref{excluded}). Also note
that the final sum over the length of the barrier is restricted to
$h \leq h_M = min (h_B ,j-j_B -1)$ with $j\geq j_B +1$,
corresponding to
the fact that, in general, the exercise point $(N,j)$ cannot be reached
from all points on the barrier. This completes our explanation of
(\ref{excluded}).

We are then left with computing $\aleph _{EB} ^{res}$. 
From its very definition it is not hard to see that $\aleph _{EB} ^{res}
[h]\equiv \aleph _{E} ^{res}[j_B +1+2h, j_B + 1+h]$ satisfies the
following recursion relation:
\begin{eqnarray}
\label{recursione1}
\aleph _{EB} ^{res} [0]~&=&~1\\
\label{recursione2}
\aleph _{EB} ^{res} [h]~&=&~\left (\begin{array}{c}
j_B +1 +2h\\
j_B +1 +h
\end{array}
\right )~-~\sum _{l=0} ^{h-1} \aleph _{EB} ^{res} [l] \left
(\begin{array}{c} 2(h-l)\\
h-l
\end{array}
\right ), ~~~~~h\geq 1,
\end{eqnarray}
with the sum in (\ref{recursione2}) removing contributions from
previously visited barrier points. Obviously $\aleph _{EB} ^{res}
[0]=1$ as there is a single path involving $j_B +1$ up moves connecting
the origin with the point $(j_B +1 ,j_B +1)$ on the tree.

To solve Equations (\ref{recursione1}) and
(\ref{recursione2}) we first combine the sum on the right-hand side of
(\ref{recursione2}) with the term on the left and rewrite the resulting
equation in the form of a discrete convolution:
\begin{equation}
\label{convolutionE}
\sum _{l=0} ^{h} \aleph _{EB} ^{res} [l] \left
(\begin{array}{c} 2(h-l)\\
h-l
\end{array}
\right ) =\left (\begin{array}{c}
j_B +1 +2h\\
j_B +1 +h
\end{array}
\right ),
\end{equation}
where the boundary condition, $\aleph _{EB} [0]=1$, is already included
as the $h=0$ contribution to (\ref{convolutionE}). Note that
(\ref{convolutionE}) can be solved by standard Laplace transform (or
$Z$-transform) techniques~\cite{miller}. Since in applying these
ideas to the more complicated American options we will lose the
convolution form -- the kernel will depend on $h$
and $l$ separately and not only through the difference, $h-l$ -- we
will proceed in a more general way and stay in ``configuration space"
until the very end.

We prefer to regard (\ref{convolutionE}) as a matrix
equation of the form:
\begin{equation}
\label{matrixeqEB}
{\bf L }_{EB} {\bf \Pi}_{EB} ^{res}
={\bf D}_{EB}.
\end{equation}
Here ${\bf \Pi}_{EB}$ and ${\bf D}_{EB}$ are $h_M +1$ dimensional
vectors, with components ${\bf \Pi}_{{EB},h} = \aleph _{EB} ^{res} [h]$
and ${\bf D}_{{EB},h} =
\left (\begin{array}{c} j_B +1 +2h\\
j_B +1 +h\end{array}\right)$, $h=0,1,2,...,h_M$,
and the $(h_M +1)\times (h_M +1)$ dimensional matrix, ${\bf L}_{EB}$, can
be written as,
\begin{equation}
\label{matrixLEB}
\left [{\bf L}_{EB}\right]_{h,l}~=~\left (\begin{array}{c}
2(h-l)\\
h-l\end{array}\right ) \theta \left (h-l\right ).
\end{equation}
Note that in (\ref{matrixLEB}) we have explicitly added
a $\theta$ function
($\theta (x)=1$ for $x\geq 0$ and vanishes otherwise) to stress that
${\bf L} _{EB}$ is a lower triangular matrix with unity along and zeros
above the diagonal. This simple observation allows us to rewrite
(\ref{matrixeqEB}) in the convenient form,
\begin{equation}
\label{rewriteLEB}
{\bf L}_{EB}={\bf 1} _{(h_M +1)\times (h_M +1)}+{\bf Q}_{EB}.
\end{equation}
where ${\bf Q} _{EB} $ is a nilpotent matrix of order $h_M$, ${{\bf Q}
_{EB}}^y =0$ for $y\geq h_M +1$; and
$\left [{\bf Q} _{EB}\right]_{h.l}=\left (\begin{array}{c}
2(h-l)\\h-l\end{array}\right )\theta (h-l-1)$.
The nilpotent property of ${\bf Q}_{EB}$ allows us to write down the
explicit solution for (\ref{matrixeqEB}),
\begin{equation}
\label{inversion1}
{\bf \Pi}_{EB} = \left [{\bf 1}~+~{\bf Q}_{EB}\right
]^{-1}{\bf D}_{EB}~=~
\sum _{R=0} ^{h_M} (-1)^R {\bf Q}_{EB} ^R {\bf D}_{EB},
\end{equation}
which in turn leads to the following formula for
the value of the option,
\begin{eqnarray}
\label{finale1}
\bar{V}_{EB}& = &\bar{V} _E - \bar{V}_{EB} ^{res}\\
\label{priceres}
\bar{V}_{EB} ^{res}& =&e^{-rT}\sum _{j=j_B +1} ^{j^*}
\sum_{h,l,R=0} ^{h_M (j)} (-1) ^R
\left (\begin{array}{c}{N-j_B -1 -2h}\\j-j_B -1 -h\end{array}\right )
\left [{\bf Q}_{EB}^R \right]_{h,l}\left (\begin{array}{c} j_B +1 +2l\\
j_B +1 +l\end{array}\right )\\
&\times& p_u ^j
(1-p_u)^{N-j}
\left (X-S_0 u^j d^{N-j}\right )\nonumber.
\end{eqnarray}
The lower limit, $j=j_B +1$, on the external sum in
(\ref{priceres})  excludes all paths unaffected by the
presence of the barrier; also we have explicitly indicated the
$N$ and/or
$j$ dependence of the various quantities involved; and have separated
out the contribution to $\aleph _{EB} [N,j]$ from unrestricted paths (the
first term on the right-hand side of (\ref{excluded})) which simply leads
to the value of the European put option given in (\ref{valueev}).

\begin{figure}[a]
\begin{center}
  \epsfxsize=400pt\epsffile{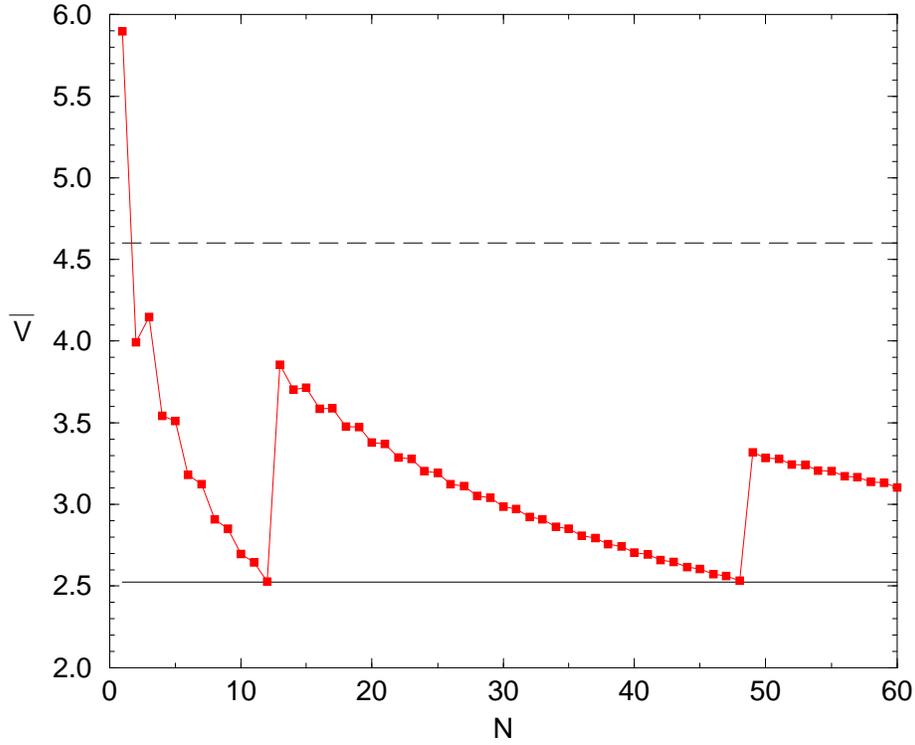}
\end{center}
\caption[a]{\protect\label{fig1}
Convergence to analytic value~\cite{rubinstein}
of a three-month European ``up-and-out" put option on
a non-dividend-paying stock as the number of binomial intervals $N$
increases. The solid line joining the squares is a guide to the eye.
The stock price $S_0$ is $60$, the risk-free interest rate $r$ is
$10 \%$, and the volatility $\sigma$ is $45 \%$. The barrier level
$H$ is at $64$. The analytic value (continuous line) is $2.524$. As
a point of reference, we give the European put option value (i.e.,
$H\rightarrow\infty$) $4.6$ after the Black-Scholes solution (dashed line). }
\end{figure}

We expect that, since we have an analytical formula, we should be able
to recover the exact solution of the continuum Black-Scholes theory
for this simplest of
barrier options~\cite{rubinstein} as was already done for conventional
European puts~\cite{cox}. Figure \ref{fig1} shows the numerical convergence
of the
binomial value of a representative ``up-and-out" European put
option to its analytic value~\cite{rubinstein}.

The same general idea used in the case of European barrier options
will now be used to write down an exact formula for the price of a
simple American option, regarding the latter as an option with an
early-exercise barrier.

\subsection{Conventional American Put Options}

Using this view to valuate American options
requires the knowledge of those points on the tree where it first becomes
profitable to exercise the option. This set of points, parametrized as
$(i, j_x [i])$, constitute the ``early exercise barrier" (EXB).
Determining the explicit form of the surface, $j_x [i]$, seems very
difficult (if at all possible) as it already implies
a knowledge of the mean value of the option at some finite number of
points on the tree. In this section we show that there is a
self-consistent exact formulation of the problem which proceeds in
the following three steps: (i) we assume that the early exercise
surface,
$j_x [i]$, is given and compute an explicit formula for the value of the
option at each point on the tree, $f(i,j; j_x [i])$, which depends
parametrically on
$j_x [i]$; (ii) the fact
that early exercise
at $(i,j)$ only occurs when $X-S_0 u^j d^{i-j} \geq f(i,j;j_x [i])$
gives us an explicit formula for the EXB which corresponds to the strict
equality,
\begin{equation}
\label{defsurface}
\left (X-S_0
u^{\tilde{j}[i;j_x [i]]} d^{i-\tilde{j}[i;j_x [i]]}
\right ) ~=~f(i,j_x [i]);~~~~~~j_x [i]=  \mbox{Int}\left
\{\tilde{j} [i;j_x [i]\right\}.
\end{equation}
[Note that, on the
right-hand side of (\ref{defsurface}) we have not used $f(i,
\tilde{j}[i])$ which might appear at first sight as a more natural
choice for defining the EXB. As will become clear below,
(\ref{defsurface}) is the simplest and most natural choice which resolves
the ambiguity of defining
$f(i,j)$ away from points on the tree.]
Finally, (iii) substituting the solution (\ref{defsurface})
into the formally
exact valuation expression gives us the value of the option.
Although this strategy leads to an exact solution of the price of
an American option, explicit numbers require rather heavy numerical
computations except in the simplest example of a straight EXB.

Let us proceed in carrying out the program outlined above
by assuming that
the EXB, i.e.,
$j_x [i]$, is explicitly given. To begin our calculation we will need
some very general properties of the barrier. These follow
from two simple characteristics of early exercise: (i) if the point
$(i,j)$ is an early exercise point, then so are all points
``deeper in-the-money", $(i,j'), ~j'=0,1,...,j-1$; and (ii) if two
adjacent points at the same time step, $(i+1, j+1)$ and $(i+1,j)$, are
both early exercise points so is the point $(i,j)$. (The latter property
follows from a conventional ``backwardation" argument~\cite{hull}
which indicates that the
average expected payoff at
$(i,j)$, discounted at the risk-free interest rate, is smaller than
the actual payoff, thus making $(i,j)$ itself an early exercise point.)
It is not hard to see that (i) and (ii) guarantee that the inner part of
the early exercise region cannot be reached without crossing the EXB.
Thus, if we define
$i_A$ to be the first time for which early exercise becomes possible
and parametrize the points on the EXB as
$(i=i_A +h,j_x [i_A +h])$ with
$h=0,1,2,...,N-i_A$, it then follows that $j_x [i_A ] = 0$.
Moreover, the structure of the tree ensures that $j_x [i]$ is a
non decreasing function of $i$; more precisely, for each time step, $j_x
[i]$ either increases by one or remains the same.

The formal expression for the price of an American
option can be written down once one recognizes that
once a path hits the EXB the option expires and thus any point on
the barrier can be reached at most once. As a result, the value of the
option is a sum of (appropriately discounted) payoffs along the barrier,
weighted by the probability of reaching each point on the barrier without
having visited the barrier at previous times.
We can then write the expected value of an American
option as:
\begin{equation}
\label{valueav}
\bar{V}_A=\sum _{h=0} ^{N-i_A} e^{-r(i_A +h)\tau} \aleph _A ^{res}[h]
p_u ^{j_x [i_A +h]} (1-p_u )^{i_A + h - j_x [i_A + h ]}\left (X-S_0
u^{j_x [i_A +h]} d^{i_A +h -j_x [i_A +h ]}\right ),
\end{equation}
where $\aleph _A ^{res}$ denotes the number of paths reaching the EXB
in $i_A +h$ time steps without having previously visited any points on
the barrier.

The counting problem can be solved along similar lines to those
followed in the case of European options:
$\aleph _A ^{res} [h]$ satisfies
an equation analogous to (\ref{recursione2}), namely,
\begin{eqnarray}
\label{recursiona1}
\aleph _{A} ^{res} [0]~&=&~1\\
\label{recursiona2}
\aleph _{A} ^{res} [h]~&=&~\left (\begin{array}{c}
i_A+h\\
j_x [i_A +h]
\end{array}
\right )~-~\sum _{l=0} ^{h-1} \aleph _{A} ^{res} [l] \left
(\begin{array}{c} (h-l)\\
j_x [i_A+h]-j_x [i_A+l]
\end{array}
\right ), ~~~~~h\geq 1,
\end{eqnarray}
where the first term on the right-hand side counts the total number of
unrestricted paths from the origin to the point $(i_A +h, j_x [i_A +h])$
on the barrier, while the second term excludes those paths which, before
reaching $(i_A +h, j_x [i_A +h])$ visited any of the previous
barrier points,
$(i_A +l, j_x [i_A +l]), ~l=0,1,2,...,h-1$~\cite{distance}.

As in the case of the European barrier option
(\ref{recursiona2}) is rewritten as a matrix equation:
\begin{equation}
\label{matrixeq}
{\bf L}_A {\bf \Pi}_A ^{res}={\bf D}_A.
\end{equation}
Here ${\bf \Pi}_A$ and ${\bf D}_A$ are $N-i_A +1$ dimensional vectors,
with components ${\bf \Pi}_{A,h} = \aleph _A ^{res} [h]$ and
${\bf D}_{A,h} =
\left (\begin{array}{c} i_A +h\\
j_x [i_A +h]\end{array}\right)$, $h=0,1,2,...,N-i_A$,
and the $(N-i_A +1 )\times (N -i_A +1)$ dimensional matrix, ${\bf L}_A$,
takes the form,
\begin{equation}
\label{matrixL}
\left[ {\bf L}_A \right]_{h,l}=\left (\begin{array}{c}
h-l\\
j_x [i_A +h]-j_x [i_A +l]\end{array}\right ),
~~~l\leq h = 0,1,2,...,N-i_A .
\end{equation}
Note that, in contrast to (\ref{matrixLEB}) and (\ref{rewriteLEB}), ${\bf
L}_A$ depends on the indices $h$ and $l$ separately; also, we have
used the identities
$j_x [i_A] =0$ and
$\left(\begin{array}{c}
i_A\\
j_x [i_A ]\end{array}\right )=1$, to incorporate the boundary condition,
$\aleph _A ^{res}[0]=1$, in (\ref{matrixeq}) in a symmetric way.
As in (\ref{rewriteLEB}), we can decompose ${\bf L}_A$ as,
\begin{eqnarray}
\label{rewriteL}
\left [{\bf L}_A\right]_{h,l}~&=&~\delta _{h,l} +\left [{\bf Q}_A
\right ]_{h,l}\\
\label{defQ}
\left [{\bf Q}_A
\right ]_{h,l}~&=&~\left (\begin{array}{c} h-l\\
j_x [i_A +h] -j_x [i_A +l]\end{array}\right )\theta \left (h-l-1\right),
\end{eqnarray}
where ${\bf Q}_A$ has nonzero elements starting just below the diagonal
and it is thus a nilpotent matrix of degree
$N-i_A$ (i.e., ${\bf Q}_A ^{N-i_A +1} = 0$). Thus,
\begin{equation}
\label{inversion}
{\bf \Pi}_A = \left [{\bf 1}~+~{\bf Q}_A\right
]^{-1}{\bf D}_A~=~
\sum _{m=0} ^{N-i_A} (-1)^m {\bf Q}_A ^m {\bf D}_A,
\end{equation}
leading in turn to the final formula for the value of the option,
\begin{eqnarray}
\label{valueavfin}
\bar{V}_A=&\sum _{h,l,m=0} ^{N-i_A}& e^{-r(i_A +h)\tau} (-1)^m \left
[{\bf Q}_A ^m\right ]_{h,l}\left (\begin{array}{c} i_A +l\\
j_x [i_A +l ]\end{array}\right )\nonumber\\&\times& p_u ^{j_x [i_A +h]}
(1-p_u )^{i_A + h - j_x [i_A + h ]}\left (X-S_0 u^{j_x [i_A +h]} d^{i_A
+h -j_x [i_A +h ]}\right ).
\end{eqnarray}

One last step is the determination of
$f(i,j)$, the value of the American put at every point $(i,j)$ on the
tree which, in turn, will allow us to derive the equation for the EXB.
This is easily done by simply translating the origin in
(\ref{valueavfin}):
\begin{eqnarray}
\label{valij}
f(i,j) &=& \sum _{{h,l,m}=i-i_A} ^{N-i_A} e^{-r(i_A +h-i)\tau}
(-1)^m \left [{\bf Q}_A ^m\right ]_{h,l}\left(
\begin{array}{c} i_A +l-i\\
j_x [i_A +l ]-j
\end{array}
\right) \nonumber \\
&\times& p_u ^{j_x [i_A +h]-j}
(1-p_u )^{i_A + h - j_x [i_A + h ]-i+j}\left (X-S_0 u^{j_x [i_A +h]}
d^{i_A +h -j_x [i_A +h ]}\right ).
\end{eqnarray}
Together with (\ref{defsurface}) this then leads to the rather
formidable-looking equation for the barrier height $\tilde{j}
[i_A +k]$ at the $(i_A +k)$-th time step ($k=0,1,...,N-i_A$), as a
functional of the barrier position at all {\em future} time steps before
expiration:
\begin{eqnarray}
\label{EXB}
\left ( X-S_0 u^{\tilde{j} [i_A +k]}
d^{i_A +k -\tilde{j} [i_A +k ]}\right ) ~&=&~\sum _{{h,l,m}=k}
^{N-i_A} e^{-r(h-k)\tau}  (-1)^m \left [{\bf Q}_A ^m\right ]_{h,l}\left
(\begin{array}{c} l-k\\ j_x [i_A +l ]- j_x [i_A +k ]\end{array}
\right )\nonumber\\&\times&  p_u ^{j_x [i_A +h]-j_x [i_A +k]}
(1-p_u )^{ h-k - j_x [i_A + h ]+j_x [i_A +k]}\\&\times&\left (X-S_0
u^{j_x [i_A +h]} d^{i_A +h -j_x [i_A +h ]}\right )\nonumber\\
\label{EXB2}
j_x [i_A +k]~~&=&~ \mbox{Int}\left \{\tilde{j} [i_A +k ]\right \}.
\end{eqnarray}
[It should now be clear that in (\ref{valij}) $j$ must be
restricted to points on the tree as the binomial coefficient
$\left (\begin{array}{c} 0\\j_x [x_A +l] -\tilde{j} [x_A +l]
\end{array}\right )$ would be ill-defined -- hence the choice
(\ref{defsurface}).]
Equations (\ref{EXB}) and (\ref{EXB2}) for the boundary
together with the formula for the value of the option,
(\ref{valueavfin}), constitute an exact pricing strategy for a
conventional American put. A similar formula for an American put with an
``up-and-out" barrier will be discussed in a future publication.

It is instructive to consider Equations
(\ref{valueavfin}), (\ref{EXB}) and (\ref{EXB2}) in the explicitly
solvable case of a straight barrier. We begin
with the observation that, at expiration,
$k=N-i_A$, (\ref{EXB}) reduces to the equation for $j^*
=\mbox{Int}[\ln(X/S_0 d^N)/\ln(u/d)]$, already defined in the case of the
European option, and thus, the barrier goes through the point $(N,j^*)$.
Moreover, starting from the exact point $(N,j^*)$ on the barrier and
decreasing $j_x [i]$ by one with each backward time step we reach $i_A =
N-j^*$ along the straight line, $j_x [i]=i-N+j^*$. Recall that, since
with each increasing time step,
$j_x [i]$ either increases by one or remains the same, this straight
line represents a lower bound for the early exercise barrier.

For this straight barrier (\ref{inversion}) and (\ref{valueavfin}) reduce
to,
\begin{eqnarray}
\label{valuestr}
P_A^{straight}~&=&~e^{-r(N-j^{*})\tau} \left (1-p_u \right )^{N-j^{*}}
\left (X-S_0 d^{N-j^{*}}\right )\nonumber\\~&+&~\sum _{h=1} ^{j^{*}}
e^{-r(N-j^{*}+h)\tau} \aleph _A ^{straight} [h]~p_u ^h
\left (1-p_u \right )^{N-j^{*}} \left (X-S_0 u^h d^{N-j^{*}}\right )
\end{eqnarray}
with
\begin{equation}
\label{countingstr}
\aleph _A ^{straight}~=~\left (\begin{array}{c} {N-j^{*} +h}\\
h\end{array}
\right )~-~\left(\begin{array}{c} {N-j^{*} +h -1}\\
h-1\end{array}
\right )
\end{equation}

We expect that the result for the true barrier should approach the
straight line formula for coarse enough time steps, $\tau > N-j^* -i_A$,
(where this $i_A$ is the first time of early exercise in the
limit of continuous-time trading).

\section{Conclusion}

We have presented a scheme for pricing options with and without barriers
on binomial trees. To the best of our knowledge ours is the
first explicit derivation of exact formulas treating barriers on binomial
trees. It is our expectation that in the
limit of continuous-time trading we
should be able to recover the few exact results available in the
literature, especially for American options\cite{kim,gao}. We also hope
that our explicit formulas may provide a framework for improving the
efficiency of numerical computations.

\section{Acknowledgements}

The authors dedicate this paper to Professor
Ferdinando Mancini, a remarkable teacher, colleague and friend, on the
occasion of his 60th birthday. We are grateful to Stanko Barle for reading
the manuscript and bringing the work of references \cite{kim} and
\cite{gao} to our attention. Finally, we acknowledge the hospitality
of the NYU Physics Department where most of this work was conceived.

\begin{thebibliography}{99}
\bibitem{distance} We recall that the number of paths between two
arbitrary points on the tree, say $(i,j)$ and $(i',j')$ (with $i'>i, \:
j'>j$) is given by the binomial coefficient, $(i'-i)!/[(j'-j)!(i'-i-j'+j)!]$.
\end{thebibliography}
\end{document}